\documentclass[prl,showpacs,preprintnumbers,amsmath,aps,twocolumn]{revtex4}

\usepackage{graphicx}
\usepackage{times}
\usepackage{xypic}
\usepackage{amsmath}
\usepackage{amsfonts}
\usepackage{amssymb}
\usepackage{amscd}
\usepackage{color}

\def\duzomniejsze{<\kern-.7mm<}
\def\duzowieksze{>\kern-.7mm>}

\def\textbf#1{{\bf #1}}
\def\be{\begin{equation}}
\def\ee{\end{equation}}
\def\bea{\begin{eqnarray}}
\def\eea{\end{eqnarray}}
\def\bse{\begin{subequations}}
\def\ese{\end{subequations}}
\newcommand{\bei}{\begin{itemize}}
\newcommand{\eei}{\end{itemize}}
\newcommand{\bee}{\begin{enumerate}}
\newcommand{\eee}{\end{enumerate}}

\def\>{\rangle}
\def\<{\langle}

\def\dt#1{{{\kern -.0mm\rm d}}#1\,}

\newtheorem{proposition}{Proposition}

\begin{document}

\title{Entanglement combing}
\author{Dong Yang$^{1,2}$}
\author{Jens Eisert$^{1}$}
\affiliation{$^{1}$Institute of Physics and Astronomy, University of Potsdam, 14476 Potsdam, Germany}
\affiliation{$^{2}$Laboratory for Quantum Information, China Jiliang University,
Hangzhou, Zhejiang 310018, China}

\date{\today}

\begin{abstract}
We show that all multi-partite pure states can, under local
operations, be transformed into bi-partite pairwise entangled
states in a ``lossless fashion'': An arbitrary distinguished party
will keep pairwise entanglement with all other parties after the
asymptotic protocol---decorrelating all other parties from each
other---in a way that the degree of entanglement of this party
with respect to the rest will remain entirely unchanged. The set
of possible entanglement distributions of bi-partite pairs is also
classified. Finally, we point out several applications of this
protocol as a useful primitive in quantum information theory.
\end{abstract}


\maketitle

In what way is multi-particle entanglement different from bi-partite one? Instances of this question
have featured prominently in the quantum information literature, motivated by the central
role entanglement plays in quantum information theory \cite{Reviews}.
Yet, in many ways,  the understanding of  multi-particle entanglement and its applications
is still unsatisfactory: Quite pragmatically speaking, while many quantum
communication and cryptographic protocols have been identified between two separated
laboratories, fewer practical protocols, say, in key distribution, are
known that directly rely on genuinely multi-partite correlations. Then, progress
on the ``traditional questions'' on multi-particle entanglement seems to have
slowed down, such as the
problem what ingredients are eventually
needed to prepare an arbitrary state
(meant in a local, asymptotically reversible fashion). What is more, it still seems not quite
clear what the exact role of multi-partite entanglement is in the known
communication protocols, and even---quite prominently---in quantum computation.
All this motivates the question in what sense one can think of multi-partite correlations
as being different from bi-partite ones, or more specifically, in what sense the former
can just be translated into the latter.

In this work we will introduce a protocol for transforming arbitrary
multi-particle entanglement into a simple, in fact, bi-partite normal form.
This protocol, referred to as entanglement combing, shows in what sense
bi-partite correlations are contained in any state, and can be viewed as a primitive
in quantum information that can be used to construct new protocols, a perspective that we outline.

The indeed surprising feature of this primitive is that this
transformation can be done in a lossless fashion: One can
simply de-correlate multi-partite entanglement always into
bi-partite one, without losing any of the entanglement between the
party holding the bi-partite entanglement and the rest. We will
first discuss the protocol, as usual under asymptotic local
operations and classical communication (LOCC). Then, we fully
classify the region of entanglement distribution that can be
achieved in the combing process. Finally, we will outline a number
of possible applications of the protocol.

\begin{figure}
\label{fig:mul} \centering
\includegraphics[scale=0.35]{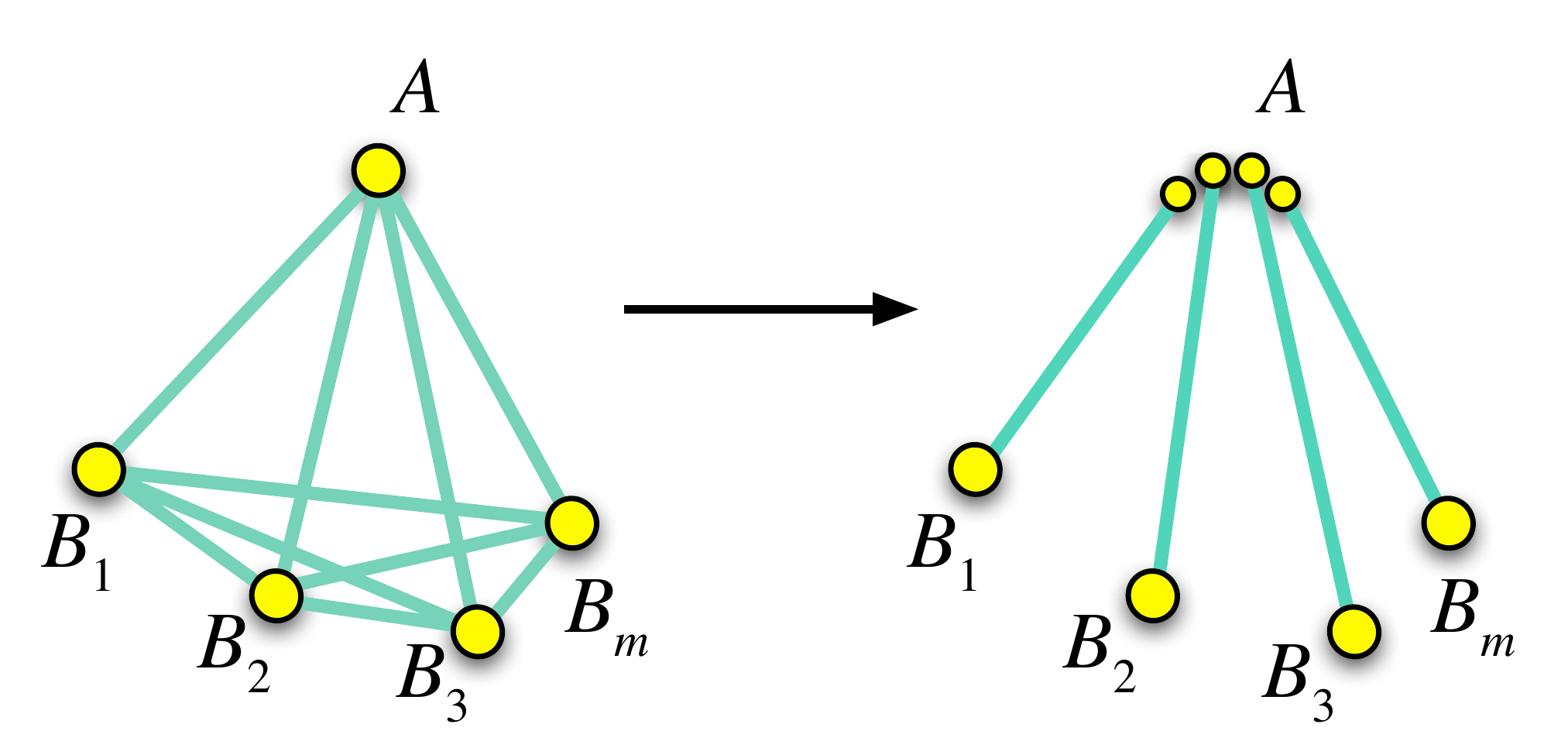}
\caption{Entanglement combing: An arbitrary multi-partite entangled
pure state
$|\phi\>_{A , B_1 ,\cdots , B_m}$ can be asymptotically
deterministically transformed into a tensor product of bi-partite
states $|\phi_1\>_{A_1, B_1}\otimes \cdots\otimes|\phi_m\>_{A_m,
B_m}$ under LOCC operations, in a way such that the bi-partite
entanglement between $A$ on the one hand and  $B_1,\dots, B_m$ on
the other hand is  preserved.}
\end{figure}

{\it The task. --} Consider an arbitrary
pure $m+1$-partite state $|\phi\>_{A, B_1, \cdots ,B_m}$ (of finite dimension)
shared among an arbitrary distinguished party (Alice) and
the other parties
(here many Bobs). Obviously, in any such state the multi-partite entanglement
structure can be very intricate. The goal is to distill tensor products
$|\phi_1\>_{A_1, B_1}\otimes \cdots\otimes|\phi_m\>_{A_m, B_m}$
of bi-partite entangled states with respect to
Alice and many Bobs under LOCC, see Fig.\ 1.
This protocol hence complements recently studied protocols for multi-partite states: One is
entanglement of assistance \cite{EA} and the other is random distillation \cite{random}.
Entanglement of assistance asks how much entanglement between two specified parties can be distilled for a pure
$m$-partite state under helpful LOCC operations of the other $m-2$ parties \cite{EA}. Random distillation in turn
asks how much pairwise entanglement can be obtained by LOCC whichever two parties
would share the final entanglement. Here we show that in fact the entanglement between a fixed party with the rest
can actually be divided into pure bi-partite states shared between the fixed party and the rest ones individually.
What is more, the final bi-partite entanglement content can be taken to be same.
It should be emphasized that as in any protocol discussing rates of entanglement transformations,
all results in this work are meant in the asymptotic setting. As usual, we simply write
$|\psi\rangle^{\otimes s}
    \rightarrow
    |\phi\rangle^{\otimes r}$ for some $r,s\geq 0$
if there is a sequence $\{i_n\}$ of integers
such that
$|\psi\rangle^{\otimes n}
    \rightarrow
    |\phi_n\rangle$
under LOCC and $\lim_{n\rightarrow\infty}
    \||\phi\rangle^{\otimes i_n}-|\phi_n\rangle \|=0$ with
    $\limsup_{n\rightarrow\infty}
    {i_n}/{n}={r}/{s}$. We will now carefully state the first main conclusion:

\begin{proposition}[Entanglement combing]
Any pure state shared between $m+1$ parties $A,B_1,\dots, B_m$
can be locally transformed, ``combed'', into a tensor product of
bi-partite states shared between $A$ and $B_1,\dots, B_m$, i.e.,
\begin{equation}
    |\phi\>_{A, B_1, \cdots ,B_m}\rightarrow |\phi_1\>_{A_1, B_1}\otimes \cdots\otimes|\phi_m\>_{A_m, B_m}
\end{equation}
under LOCC. This can be done in a way such that the entanglement of
$A$ with respect to $B_1,\dots, B_m$ is left unchanged,
$\sum_k E_k = \sum_k S(\rho_{A_k}) = S(A)$.
\end{proposition}

Here, $S(A)$ is the initial von Neumann entropy of $A$, and the entanglement of a
bi-partite pure state is measured as the usual entanglement entropy
\begin{equation}\nonumber
    E(|\phi_k\rangle_{A_k, B_k} ):=S(\rho_{A_k})=:E_k,
\end{equation}
$\rho_{A_k}$ denoting the respective reduced state. In the proof
of this statement -- delayed to the proof of possible
distributions -- two techniques feature strongly: One is the
protocol quantum state merging \cite{HOW} and the other is a Lemma
\cite{HOW} that implies the entanglement of assistance. In a way,
all technicalities when identifying the reachable set are related
to appropriately generating appropriate resources using these
protocols in substeps, then using them in later steps, to again
create suitable resources and so on, subtly balancing trade offs,
in a way that yields asymptotically the correct rates.

{\it T1 (Quantum state merging):} For a pure tripartite state $|\phi\>_{A,B,C}$, the entanglement
cost of merging $A$ to $B$ under the reference
$C$ is equal to the conditional entropy $S(A|B)=S(AB)-S(B)$. When $S(A|B)$ is positive, $S(A|B)$
entanglement has to be consumed to perform merging. When $S(A|B)$ is negative, merging can be performed under LOCC, and moreover $-S(A|B)$ entanglement is obtained.

{\it T2 (Assisting):} For a pure tripartite state
$|\phi\>_{A,B,C}$, if $S(B)>S(A)$, then there exists a complete
measurement on $C$ such that the resulting state of $ABC$ is the
ensemble $\{p_k,|\psi^k \>_{A,B}\otimes |k\>_C\}$ satisfying
$S(\rho_A^k)\approx S(A)$.

{\it Distribution of entangled pairs. --}
Now we know that entanglement between an arbitrary
distinguished party
Alice and all other parties as a whole can be ``combed'' under LOCC into
pairwise entanglement such that the sum of the pairwise entanglement is just the initial entanglement.
Clearly, there is no reason to assume that this final distribution is unique: This very distribution is, however,
important when thinking of new protocols based on this primitive. We now turn to giving a complete answer to
the possible distributions of entangled pairs, reminding of the quantum marginal problem \cite{Marginals}:

\begin{proposition}[Distribution of entangled pairs]
    The feasible set of different
    entanglement distributions in entanglement combing $F= \{(E_1,E_2,\cdots,E_m)\}$
    for a given initial state
    $|\phi\>_{A, B_1, \cdots ,B_m}$ is a polytope: It is the
    positive part of the polytope the extreme points of
    which being given by merging the states of $m$ parties to $A$ in different orders.
\end{proposition}

{\it Proof.} It is clear that, if such a protocol exists, $\sum_k
E(|\phi_k\rangle_{A_k, B_k} )\leq S(\rho_{A})$, as the degree of
entanglement between Alice and the multiple Bobs cannot increase
\cite{comp-tele}. The surprising fact is that the upper bound can
indeed be achieved. Let us first briefly see that such a protocol
exists (although this protocol not being constructive). Suppose we
consider to deal with $B_m$ firstly. If $S(A)\ge S(B_1\cdots
B_{m-1})$, then we perform $T1$ that will merge $B_m$ into $A$ and
additionally $-S(B_m|A)=S(A)-S(B_m A)=S(A)-S(B_1 \cdots B_{m-1})$
of entanglement between $A$ and $B_m$ will be obtained as an
asymptotic rate, where we use the fact that for a pure state
$|\phi\rangle_{X,Y}$, $S(X)=S(Y)$. That is,
$|\phi\>_{A,B_1,\cdots, B_{m}}\rightarrow |\psi\>_{(B_m,
A),B_1,\cdots, B_{m-1}}\otimes |\xi\>_{A_m, B_m}$ such that $S(B_m
A)+E( |\xi\>_{A_m, B_m})=S(A)$, $S(A)$ denoting the initial local
entropy of $A$. If $S(A)< S(B_1\cdots B_{m-1})$, then we perform
$T2$ to achieve the ensemble $\{p_k, |\phi\>_{A,B_1,\cdots
,B_{m-1}}^k\otimes |k\>_{B_m}\}$ such that $S(\rho_A^k)\approx
S(A)$. In both cases the entropy of the $A$ remains invariant up
to asymptotically negligible corrections, and $B_m$ is decoupled.
However, now we are left with a $m$-partite state among $A$ and
$B_1,\cdots, B_{m-1}$.  Next we deal with $B_{m-1}$ and iterate
the strategy until we obtain the final state of the form
$|\phi_1\>_{A_1, B_1}\otimes \cdots\otimes|\phi_m\>_{A_m, B_m}$.
During each step the entropy of $A$ remains invariant, again up to
corrections not relevant for the rate.

We now turn to the actual proof of the possible distributions.
There are two steps of the argument to arrive at the conclusion.
In the first step, we formulate a convex outer approximation $F'
\supset F$ of the set, noting that we get better rates if we allow
negative quantity of entanglement shared between Alice and the
Bobs. A negative value means that entanglement is actually
consumed instead of being obtained at the final stage, or in other
words entanglement should be borrowed in order to accomplish the
task. If negative values are allowed, the combing can be regarded
as merging process and the extreme points of the convex set $F' $
are obtained by merging the states of all Bobs except the last one
to that of Alice in different orders. Convexity of $F'$ is readily
shown by the time-sharing technique \cite{CT}.
For the $m+1$-partite state, one point $(E_1,\dots,E_m)$ is
obtained by the merging order: Say, firstly merging $B_m$ to $A$,
secondly $B_{m-1}$ to $AB_m$, thirdly $B_{m-2}$ to $AB_{m-1}B_m$
and so on. So we get $E_1=S(B_1)$, $E_2=S(AB_3 \dots B_m)-S(B_1)$,
until $E_{m-1}=S(AB_m)-S(B_1\dots B_{m-2})$, $E_m=S(A)- S(B_1\dots
B_{m-1})$, evidently summing to $S(A)$. These $m!$ points are the
extreme points of $F'$: The reason comes from quantum distributed
compression. Imagine that if after the merging protocol Bobs
compress their parts and send to a new party, say $Z$, then $Z$ is
capable to recover the original state $\rho_{B_1,\dots,B_m}$ while
preserving the coherence with Alice. $(E_1,\dots,E_m)$ is an
extreme point in the distributed compression \cite{HOW}: First
compressing and sending $B_1$, then $B_2$, \dots, $B_m$ in a
sequence. All other extreme points are found similarly, and $F'$
is a polytope. $F\subset F'$ or a contradiction will arise.

In the second step, we show that the combing region is just the
intersection of this polytope with the positive cone: That is,
each non-negative point can be achieved without borrowing
entanglement beforehand. At the final stage of combing, obviously
only non-negative quantities of entanglement are allowed. We know
how to achieve any point in $F'$ with borrowing, and know that $F$
must contain only positive points, hence we are left to show that
there exists a non-borrowed protocol approximating all
non-negative points arbitrarily well. We will use the techniques
of ``breeding'' in entanglement distillation \cite{err-cor} and
time-sharing in information theory \cite{CT}. Moreover, it will be
a sequential scheme labeled by rounds $r$, where each is an
asymptotic protocol in its own right. The entire procedure is
hence meant as a sequence of protocols on more and more input
copies, where the rates in the asymptotic versions of each round
are preserved. The main idea is to prepare just the right
resources for the next round, and amplify the output and find that
initially borrowed resources become asymptotically negligible.

Let us consider any point $V\in F$ in its interior. Using
Caratheodory's theorem, we know that $V$ can be written as a
convex combination of no more than $m+1$ extreme points of the
polytope, labeled $P,Q,\dots, S$, $V=p P + q Q + \dots + s S$,
which is pointwise strictly positive by assumption \cite{Note}. Let us denote
with $P^+$ the positive part of $P$ and with $P^-$ the negative
part, and similarly for $Q,\dots, S$. Let us denote with
$|+\rangle_{A,B_k}$ a maximally entangled qubit pair between $A$
and $B_k$.

In the first round $r=1$, we will consider the (asymptotic
protocol) that performs entanglement assistance on some number of
initial copies of $|\phi\rangle_{A,B_1,\dots, B_m}$ in order to
prepare the integer number $\lfloor n_1\rfloor$ of maximally
entangled pairs $|+\rangle_{A,B_1}$ between $A$ and $B_1$, of with
$n$ better and better approximation, where
\begin{equation}
    n_1:= n (pP^-_1 + q Q^-_1+\dots s S^-_1).
\end{equation}
$n$ will then be the quantifier of the asymptotic limit of the protocol, and analogously
for parts $2,\dots, m$.
This process, which may be inefficient, then yields
$\lfloor n_1\rfloor $ specimens of $|+\rangle_{A,B_1}$ shared between $A$ and $B_1$, $\lfloor n_2\rfloor$ of
$|+\rangle_{A,B_2}$ between  $A$ and $B_2$, asymptotically perfectly, with arbitrarily small norm
error in each round, and so on.

For the second round, $r=2$, we now know that from the protocols
at $P,Q,\cdots,S$ under borrowing, and the technique of time
sharing, grouping the prepared bipartite entanglement,
using asymptotic reversibility of pure-state bi-partite state transformations,
\begin{equation}
    |\phi\rangle_{A,B_1,\dots, B_m}^{\otimes n}
    |+\rangle_{A,B_1}^{\otimes n_1}\dots
    |+\rangle_{A,B_m}^{\otimes n_m}
    \rightarrow |+\rangle_{A,B_1}^{\otimes k_1}
    \dots
    \rightarrow |+\rangle_{A,B_m}^{\otimes k_m}
\end{equation}
holds as an asymptotic transformation, where
$ k_j:= n (pP^+_j + q Q^+_j+\dots s S^+_j)$,
for $j=1,\dots, m$. This can be reached by performing the borrowing merging
protocol $P$ with a relative weight of $p$, then $Q$ with a relative weight of $q$, until
$S$ with a relative weight of $s$, and then combing the resulting maximally entangled
pairs appropriately. This is possible, as the resources needed in the borrowing
are available. Define now $x_j:= k_j/n_j$, as the amplification ratio.  By definition,
$x_j> 1$ for all $j$; due to positivity, there
will be more entangled pairs available after this step at any position.
Hence, $\lfloor  k_1 \rfloor$ specimens of $|+\rangle_{A,B_1}$ will
be available after this step, asymptotically perfectly, and
similarly for the other parties.

For the third step, $r=3$, define $x:=\min\{x_j:j=1,\dots, m\}>1$.
Now one again borrows maximally entangled pairs to assist the next
step: We will use $\lfloor n x\rfloor$ copies of maximally
entangled pairs to perform $P$ again on $\lfloor n p x \rfloor$
copies, $Q$ on $\lfloor n q x \rfloor$ copies, until $S$ on
$\lfloor n s x \rfloor$ copies. This in turn is used in the next
steps $r$. At large $r$ we calculate the relative weight of the initially
consumed $n n_0$ copies from assisting.
The total number of consumed copies in $r$ rounds is then
$nn_0+\sum_{i=0}^{r}nx^i=n(n_0+{(x^{r+1}-1)}/{(x-1)})$.
Since
$x>1$, the initially consumed copies from assisting will have a
logarithmic weight in $r$ asymptotically in $r$ that is negligible
at large $r$. The entire asymptotic protocol amounts to taking the
$r, n\rightarrow \infty$ limit, in that the appropriate rate and the
norm approximation can be achieved to arbitrary accuracy.
In the end we can obtain the rate
at the interior point $V\in F$ without borrowing.

Notice that for the protocol to continue it is required that $x>
1$. If $x<1$, less and less entanglement is gained at one position
such that less and less copies can be activated further. The
condition that the activation can be amplified is just the
requirement that $V$ lies in the positive part of $F$. Now, if we
are at a boundary point of $F$, at a face of the polytope, one can
approximate $V$ with a sequence of efficient protocols arbitrarily
well, and the actual set of asymptotically reachable points is
closed. Notably the argument established here can also be used in
other protocols with borrowed resources \cite{Remarks}.

{\it Applications. --}
Once we have obtained the region of entanglement distribution, we will now turn to sketching
potential applications of this protocol in quantum information theory.

{\it 1. Distributed compression.}
Multi-partite entangled states can be employed as a resource in quantum distributed compression. From Schumacher compression \cite{schmu-compression}, it is known that a source emitting states with $\rho$ can be compressed into a Hilbert space of dimension
$S(\rho)$ transmitted, such that the original data can be decoded faithfully. In quantum distributed compression,
quantum data are distributed among many Bobs who are required to separately compress 
their data and send their parts to a common party
Alice who can decode the whole data faithfully. It has just recently been proven \cite{HOW} that the qubits that are required to transmit is still $S(\rho)$ though the classical scenario was known for a long time \cite{distr-compression}. Notice that the 
compressed data are transmitted either through ideal channels or teleported via ebits shared 
between Bobs and Alice. The entanglement combing provides a way how the parties can 
employ their shared multi-partite state as resource to complete the task. The multi-partite 
can be used to replace the ideal quantum channels and the bi-partite entangled states. 
The whole protocol works like this: First we apply the combing entanglement to obtain 
bi-partite entanglement between Alice and many Bobs.
Then we apply distributed compression to compressing the quantum data. Finally,
we teleport the compressed data \cite{teleportation}. The region of distributed compression 
and that of the combing are therefore
both known. If there exists an overlap between these regions,  the compressed data can 
be transmitted by the state.

{\it 2. New criteria for multi-partite LOCC transformations. }
Entanglement combing provides a lower bound for the rate of
multi-partite states transformation under LOCC operations. The
entanglement of a multi-partite state can be combing at any party.
For a pure $(m+1)$-partite states we actually have $m+1$ different
regions for different combing processes. Consider two
$m+1$-partite states $|\phi\>_{A,B_1,\cdots ,B_m}$ and
$|\psi\>_{A,B_1,\cdots ,B_m}$. If
$r(S(\psi_1),S(\psi_2),\cdots,S(\psi_m))$ lies in the region $F$
of the combing protocol of $|\phi\>_{A,B_1,\cdots ,B_m}$, then a
single copy of $|\phi\>_{A,B_1,\cdots ,B_m}$ can  asymptotically
be transformed into $r$ copies of $|\psi\>_{A,B_1,\cdots ,B_m}$
under LOCC that immediately gives a lower bound for the rate,
$\psi_k$ denoting reduced states. The transformation process is:
First we perform the combing protocol on $|\phi\>_{A,B_1,\cdots
,B_m}$ to obtain the bi-partite entangled states between, then
Alice prepares the multi-partite state $|\psi\>_{A,B_1,\cdots
,B_m}$ and compresses different parts of $B_k$ by Schumacher
compression, and then teleports the compressed data of $B_k$ to
different Bobs. After received the data, the Bobs decode the data
such that $|\psi\>_{A,B_1,\cdots, B_m}$ appears among the parties.
Notice we can choose any party as Alice, so a certain choice leads
to the optimal bound.

{\it 3. Quantifying the multi-partite character of entanglement. }
The intuition is that there should exist nontrivial bipartite
entanglement distribution in a genuine multipartite entangled
state. We know that the region is convex set in a hyperplane in
high dimension space. The geometry of the region of entanglement
distribution could provide the information of genuine
multi-partite entanglement. A simple example is that if the state
$|\psi\>_{A,B_1,\cdots, B_m}$ is of the form
$|\phi\>_{A,B_1,\cdots ,B_k}\otimes|\psi\>_{A,B_{(k+1)},\cdots,
B_m}$, then no genuine $m+1$-multi-partite entanglement should
exist. This fact is reflected in the rate region is that the
hyperplane will have lower dimension while generically it has
dimension $m-1$. A simple geometric quantity is the volume of the
polytope which we conjecture would be a potential quantity for
genuine multi-partite entanglement (but also lower-dimensional
quantities could possibly be used).

{\it 4. Relationship to the quantum marginal problem. } The
protocol reminds in several ways of the celebrated quantum
marginal problem, one way of formulating it for qubits being as
such: Given $m+1$ parties $A, B_1,\dots, B_m$ and a vector
$(s_1,\dots, s_{m+1})$ with entries from $[0,1/2]$. Is there a
pure state $|\psi\rangle_{A,B_1,\dots, B_m}$ such that the spectra
of the local reductions of $A$ and $B_1$ to $B_m$ are $\{s_k,
1-s_k\}$, $k=1,\dots, m+1$? In fact, the feasible region of
possible $(s_1,\dots, s_{m+1})$ with a yes answer is a polytope
\cite{Marginals}. In general, the marginal problem ask the
question whether the given conditions are compatible. There are
two connections here: On the one hand, the possible combing
polytopes are governed by the entropies of collections of
subsystems that are consistent with a pure state. On the other
hand, one can ask similar question in entanglement combing: Given
one positive point, we easily know there exists one state on which
we comb and obtain the distribution of bipartite states
corresponding to this point. A compatibility question is then:
Given two (or several) points, whether a single pure state exists
giving rise to both points under combing.

{\it 5. Multipartite quantum communication. } Quite clearly, any multi-partite task of
quantum communication based on known resources, one can always first bring
the multi-partite state into a ``combed'' bi-partite form. Then, using the powerful machinery
of bi-partite pure state entanglement manipulation, one immediately arrives at bounds
of rates to the original protocol. In this sense, we expect this protocol also to be a helpful
tool for getting bounds to a number of multi-partite quantum communication protocols.

{\it Summary and outlook.}
In summary, we have established a
new protocol for multi-partite pure states, showing that all pure  multi-partite pure states
can be transformed into a bi-partite form, entirely preserving the bi-partite entanglement
with a party. We also identified the convex set of attainable final configurations in a quantitative
manner, giving rise to a new toolbox useful in constructing multi-partite tasks and assessing
rates for known ones, a perspective that seems quite promising when further fleshing out
the potential of multi-partite quantum information processing.

{\it Acknowledgements. --} We would like to thank M.\ Christandl,
D.\ Gross, M.\ Horodecki, and J.\ Oppenheim for stimulating
discussions. D.Y. acknowledges the support from NNSF of China
(Grant No. 10805043), J.E. support from the EU (QAP, COMPAS,
MINOS) and the EURYI.


\begin{thebibliography}{99}
\bibitem{Note}
    Note that to check whether for given extreme points there exists a
    positive point in their convex combination amounts to solving a
    linear feasibility problem.
\bibitem{Remarks}
    The entanglement combing is not a reversible process generally \cite{mres} though it is
    reduced to the entanglement concentration in the case of bi-partite states where it is known that the process
    is fully asymptotically reversible \cite{comp-tele}. The modified version of entanglement assistance is discussed in
    Ref.\ \cite{LOW} where borrowed ebits between some pairs are allowed.
\end{thebibliography}
\end{document}